\documentclass[12pt]{article}

\usepackage{amsmath}
\usepackage{amssymb}
\usepackage{amsthm}
\usepackage{mathrsfs}
\usepackage[T1]{fontenc}
\usepackage{enumerate}
\usepackage{color}

\def\eqn{equation}

\def\tfn{transformation}

\def\decomp{decomposition}

\def\sm{sigma model}
\def\pl{Poisson--Lie}

\def\dd{Drinfel'd double}

\def\4diml{four-dimensional}
\def\-1{^{-1}}

\def\coor{coordinate}

\def\e{{\rm e}}
\def\cd{{\mathfrak d}}
\def\cg{{\mathfrak g}}
\def\tcg{\tilde{{\mathfrak g}}}

\def\wt{\tilde}
\def\wh{\widehat}
\def\wwt{\widetilde}
\def\sm{sigma model}
\def\pltp{Poisson--Lie T-pluralit}
\def\pltd{Poisson--Lie T-dualit}
\def\natd{non-Abelian T-dualit}

\def\cf{{\mathcal {F}}}

\newcommand{\unit}{\mathbf{1}}
\newcommand{\M}{\mathscr{M}}
\newcommand{\D}{\mathscr{D}}
\newcommand{\G}{\mathscr{G}}
\newcommand{\tG}{\widetilde{\mathscr{G}}}
\newcommand{\hG}{\widehat{\mathscr{G}}}
\newcommand{\bG}{\bar{\mathscr{G}}}

\newcommand{\N}{\mathscr{N}}

\newcommand{\PL}{Poisson--Lie}
\newcommand{\wb}{\bar}

\begin{document}



\title{\pltp y revisited. Is T-duality unique?}
\author{Ladislav Hlavat\'y\footnote{hlavaty@fjfi.cvut.cz}\\
{\em Faculty of Nuclear Sciences and Physical Engineering,} \\Czech Technical University in Prague
\\
Ivo Petr\footnote{ivo.petr@fit.cvut.cz}\\{\em Faculty of Information Technology,} \\Czech Technical University in Prague}

\maketitle
\begin{abstract}
We investigate (non-)Abelian T-duality from the perspective of Poisson--Lie T-plurality. We show that sigma models related by duality/plurality are given not only by Manin triples obtained from decompositions of Drinfel'd double, but also by their particular embeddings, i.e. maps that relate bases of these decompositions. This allows us to get richer set of dual or plural \sm s than previously thought. That's why we ask how T-duality is defined and what should be the ``canonical'' duality or plurality transformation.


\end{abstract}
\tableofcontents


\section{Introduction}

The notion of (non-)Abelian T-duality
\cite{buscher:ssbfe, delaossa:1992vc, rocekver:duality} of sigma
models relies on the presence of symmetries of the sigma model
backgrounds. Whenever there is such a symmetry, one may gauge it to
arrive at a model related to the original one by T-duality. This technique, extended to RR fields in \cite{1012.1320 sfethomp,1104.5196LCST}, is used frequently to generate new supergravity solutions, see e.g. \cite{KlebWitt,INNSZ} and references therein.
Non-Abelian T-duality, however, does not preserve the symmetries,
and it may not be possible to return back to the initial model.
\pltd y, introduced in the seminal paper \cite{klise} by Klim\v
c\'ik and \v Severa, treats both models equally and offers a remedy
to this issue.

The algebraic structure underlying \pltd y is the \dd, a Lie group $\D$ that decomposes into two Lie subgroups
$\G$ and $\tG$ of equal dimension.  In case of (non-)Abelian
T-duality the former represents group of symmetries of the initial
sigma model, while the latter is Abelian. There are also
\dd s where both $\G$ and $\tG$ are non-Abelian. In such a case
the symmetry of the initial model is replaced by the so-called
\pl\ symmetry (or
generalized symmetry), see \cite{klim:proc}, and the full \pltd y
transformation applies. {Nevertheless, the presence of symmetries remains crucial if one wants to dualize a particular background \cite{Bouwknegt2017}. Recently (\PL) T-duality also appears as an important tool in the study of integrable models and their deformations \cite{BorWulff1, BorWulff, HoaSei}.}

Since duality exchanges roles of $\G$ and $\tG$, we may
understand it in terms of \dd\ as a switch between decompositions
$(\G|\tG)$ and $(\tG|\G)$ of $\D$. The authors of
\cite{klise} mention the fact that a \dd\ $\D$ can have other
decompositions $(\cal K|\widetilde{\cal K}),(\widetilde {\cal K}|{\cal
K}),\ldots $ beside $(\G|\tG)$ and $(\tG|\G)$. All these
decompositions can be used to construct mutually related \sm s. The
transformation of the initial model constructed by
decomposition $(\G|\tG)$ to a model constructed by $(\cal
K|\widetilde{\cal K})$ was later denoted \pltp y \cite{unge:pltp}.
Examples of \sm s related by \pltp y were studied e.g. in
\cite{unge:pltp, hlasno:3dsm2}, and decompositions of
low-dimensional \dd s were classified in
\cite{gom:ctd, hlasno:pltdm2dt, snohla:ddoubles}.

The goal of this paper is to show, using simple examples, {that \sm s
related by \pltd y/plurality are given not only by the algebraic
structure of \decomp s of Lie algebra of the \dd\ into Manin
triples, but also by the particular embedding of Manin triples, i.e.
maps that relate bases in various \decomp s.} {For this purpose we shall consider the simplest possible case of \dd\ accomodating plurality, i.e. a four-dimensional semi-Abelian \dd.

After summarizing \pltp y in section \ref{pltpuvod} we identify transformations that yield equivalent sigma model backgrounds. In section \ref{2dsm} we develop examples of dual/plural models whose geometric properties depend on the choice of matrices transforming bases of Manin triples, and, in section \ref{4dsm}, we show that nonequivalent models can be obtained even if we do not change the Manin triple at all. We study this ``\PL\ T-identity'' further in section \ref{PLbyPLids} trying to identify what the ``canonical'' duality/plurality should be.}

\section{\pltp y of \sm s}\label{pltpuvod}

Let $\M$  be $n$-dimensional (pseudo-)Riemannian target manifold and consider sigma model on $\M$  given by  Lagrangian
\begin{equation}\label{Lagrangian}
{\cal L}=\partial_- \phi^{\mu}\cf_{\mu\nu}(\phi)
\partial_+ \phi^\nu,\quad \phi^\mu=\phi^\mu(x_+,x_-),\quad \mu=1,\ldots,n
\end{equation}
where tensor $\cf=\mathcal G + \mathcal B$ defines metric and torsion potential of the target manifold. Assume that there is a  $d$-dimensional Lie group $\G$ with free action on $\M$ that leaves the tensor invariant.
The action of $\G$ is transitive on its orbits, hence we may locally consider $\M\approx (\M/\G) \times \G = \N \times \G$, and introduce adapted coordinates
\begin{equation}\label{adapted}
x^\mu=\{s_\delta,g_a\},\qquad \delta=1, \ldots, n-d,\ \ a=1, \ldots, d
\end{equation}
where $s_\delta$ label the orbits of $\G$ and are treated as spectators and $g_a$ are group coordinates \cite{hlapevoj, hlafilp:uniq}. Dualizable sigma model on $\N \times \G$ is given by
tensor field $\mathcal{F}$ defined by $n\times n$ matrix $E(s)$ as
\begin{equation}\label{F}
        \cf(s,g)=\mathcal{E}(g)\cdot E(s)\cdot \mathcal{E}^T(g), \qquad \mathcal{E}(g)=\left(\begin{array}{cc} \unit & 0 \\ 0 &e(g) \end{array}\right)
\end{equation}
where $e(g)$ is $d\times d$ matrix of components of right-invariant Maurer--Cartan form $(dg)g^{-1}$  on $\G$.

Using non-Abelian T-duality one can find dual sigma model on $\N \times \tG$, where $\tG$ is Abelian subgroup of semi-Abelian \dd\ $\D$ that splits into subgroups $\G$ and $\tG$. The necessary formulas will be given in the following subsection as a special case of \pltp y. In papers \cite{hlapet:cqgrav, php, hpp}, non-Abelian T-duals of \sm\ in flat torsionless  four-dimensional background were constructed. The groups $\G$ were then  subgroups of the Poincar\'e group \cite{PWZ}.

\subsection{\PL\ T-plurality with spectators}\label{pltpuvod1}

For certain \dd s several decompositions may exist. Suppose that $\D=(\G|\tG)$ splits into another pair of subgroups $\hG$ and $\bG$. Then we can apply the full framework of \PL\ T-plurality \cite{klise, unge:pltp} and find sigma model on $\N \times \hG$.

The $2d$-dimensional Lie algebra $\cd$ of the \dd\ $\D$ is equipped with an ad-invariant non-degenerate symmetric bilinear form $\langle . , . \rangle$.
Let $\mathfrak{d}=\mathfrak g \oplus \wt{\mathfrak g}$ and $\mathfrak{d}=\wh{\mathfrak{g}} \oplus \wb{\mathfrak{g}}$ be two decompositions (Manin triples $(\mathfrak{d},\mathfrak g , \wt{\mathfrak g})$ and $(\mathfrak{d},\wh{\mathfrak{g}}, \wb{\mathfrak{g}})$) of $\cd$ into subalgebras that are maximally isotropic with respect to $\langle . , . \rangle$. The pairs of mutually dual bases $T_a \in \mathfrak g,\ \widetilde{T}^a \in \wt{\mathfrak g}$ and $\wh T_a\in\wh{\mathfrak{g}}, \ \wb{T}^a \in \wb{\mathfrak{g}}$, $a=1, \ldots, d,$ satisfying
\begin{align}
\label{dual_bases}
\langle T_a, T_b \rangle & = 0, & \langle \widetilde{T}^a, \widetilde{T}^b \rangle &= 0, & \langle T_a, \widetilde{T}^b \rangle & = \delta_a^b,\\
\nonumber \langle \wh T_a, \wh T_b \rangle & = 0, & \langle \wb{T}^a, \wb{T}^b \rangle & = 0, & \langle \wh T_a, \wb{T}^b \rangle & =  \delta_a^b
\end{align}
are related by transformation
\begin{equation}\label{C_mat}
\begin{pmatrix}\wh T \\
\wb T \end{pmatrix} = C \cdot \begin{pmatrix}T \\ \widetilde T \end{pmatrix}
\end{equation}
where $C$ is an invertible $2d\times 2d$ matrix.
Due to ad-invariance of the bilinear form $\langle . , . \rangle$ the algebraic structure of $\mathfrak{d}$ is given both by
\begin{equation}
\label{commutation_on_d}
[T_i,T_j]=f_{ij}^k T_k, \qquad [\widetilde T^i, \widetilde T^j]=\wt f^{ij}_k \widetilde T^k, \qquad [T_i,\widetilde T^j]=f_{ki}^j \widetilde T^k + \wt f_i^{jk} T_k\end{equation}
and
\begin{equation}
\label{commutation_on_d2}
[\wh T_i, \wh T_j]=\hat f_{ij}^k \wh T_k, \qquad [\wb T^i, \wb T^j]=\wb f^{ij}_k \wb T^k, \qquad [\wh T_i,\wb T^j]=\hat f_{ki}^j \wb T^k + \wb f_i^{jk} \wh T_k.
\end{equation}
Given the structure constants $F_{ij}^k$ of $\mathfrak{d}=\mathfrak g \oplus \wt{\mathfrak g}$ and $\wh F_{ij}^k$ of $\mathfrak{d}=\wh{\mathfrak{g}} \oplus \wb{\mathfrak{g}}$, the matrix $C$ has to satisfy equation\footnote{Conditions on $C$'s are more restrictive than those for ``NATD group'' investigated in \cite{LuOst} (We are grateful to D. Osten for bringing our attention to this paper) but as said in the Introduction, our main goal is to present dependence of geometrical properties of the \pl{} plural \sm s on matrices $C$.}

\begin{equation}
\nonumber
C_a^p C_b^q F_{pq}^r = \wh F_{ab}^c C_c^r.
\end{equation}
To preserve the bilinear form $\langle . , . \rangle$ and thus \eqref{dual_bases}, $C$ also has to satisfy
\begin{equation}
\nonumber
C_a^p C_b^q B_{pq}=B_{ab}
\end{equation}
where $B_{ab}$ are components of matrix $B$ that can be written in block form as
\begin{equation}
\nonumber
B=\begin{pmatrix} \mathbf{0}_d & \mathbf{1}_d \\ \mathbf{1}_d & \mathbf{0}_d \end{pmatrix}.
\end{equation}
{In other words, $C$ is an element of $O(d,d)$ but, unlike the case of Abelian T-duality, not every element of $O(d,d)$ is allowed in \eqref{C_mat}.}

For the following formulas it will be convenient to introduce $d \times d$ matrices $P, Q, R, S$ as
\begin{equation}\label{pqrs}
    \begin{pmatrix}T \\ \widetilde T \end{pmatrix}= C^{-1} \cdot \begin{pmatrix}\wh T \\
\wb T \end{pmatrix} =
   \begin{pmatrix} P & Q \\ R & S \end{pmatrix} \cdot \begin{pmatrix}\wh T \\
\wb T \end{pmatrix}.
\end{equation}
To accommodate the spectator fields we have to extend these to $n\times n$ matrices
\begin{equation}
\nonumber
\mathcal{P} =\begin{pmatrix}\unit &0 \\ 0&P \end{pmatrix}, \qquad \mathcal{Q} =\begin{pmatrix}0&0 \\ 0&Q \end{pmatrix}, \qquad \mathcal{R} =\begin{pmatrix}0&0 \\ 0&R \end{pmatrix}, \qquad \mathcal{S} =\begin{pmatrix}\unit &0 \\ 0& S \end{pmatrix}.
\end{equation}
It is also advantageous to introduce block form of $E(s)$ as
\begin{equation}
\nonumber
 E(s)=\left(\begin{array}{cc}
 E_{\alpha\beta}(s) & E_{\alpha b}(s) \\
 E_{a\beta}(s)      &E_{ab}(s)    \end{array}\right),\
 \ \alpha,\beta=1,\ldots,n-d,\ \ a,b=1,\ldots,d.
\end{equation}
The sigma model on $\N \times \hG$ related to \eqref{F} via \pltp y is given by tensor $\wh{\cf}(s,\hat g)$ that is calculated as
\begin{equation}
\label{Fhat}
\widehat{\cf}(s,\hat g)= \mathcal{\widehat E}(\hat g)\cdot \widehat E(s,\hat g)\cdot \mathcal{\widehat E}^T(\hat g), \qquad \mathcal{\widehat E}(\hat g)=
    \begin{pmatrix}
    \unit & 0 \\ 0& \wh e(\hat g)
    \end{pmatrix}
\end{equation}
where
\begin{equation}\label{Fhat2}
\wh E(s,\hat g)=(\unit+\wh E(s)\cdot \wh{\Pi}(\hat g))^{-1}\cdot \wh E(s), \qquad
  \widehat\Pi(\hat g)=\left(\begin{array}{cc}  0 & 0 \\ 0 & \widehat b(\hat g)\cdot\widehat a^{-1}(\hat g)
  \end{array}\right),
\end{equation}
matrices $\widehat b(\hat g)$ and $\widehat a(\hat g)$ are submatrices of the adjoint representation
$$ ad_{{\hat g}\-1}(\wh T) = \widehat b(\hat g)\cdot\wb T+\widehat a^{-1}(\hat g)\cdot\wh T,$$
and the matrix $\wh E(s)$ is obtained by formula
\begin{equation}\label{E0hat}
\wh E(s)=(\mathcal{P}+ E(s) \cdot \mathcal{R})^{-1} \cdot (\mathcal{Q}+E(s) \cdot \mathcal{S}).
\end{equation}
Therefore, it is necessary that
\begin{equation}\label{cond_plural}
\det\, \left(\mathcal P+E(s)\cdot \mathcal R \right)=\det\, \left(P + E_{ab}(s)\cdot R\right)\neq 0.
\end{equation}
These formulas reduce to formulas for \PL\ T-duality if we choose $P=S=\mathbf{0}_d$ and $Q=R=\mathbf{1}_d$. {Furthermore, for a semi-Abelian \dd\ the well-known Buscher rules for \natd y are restored. If there are no spectators, i.e. if $n=d$, the plurality is called atomic.}

\subsection{Equivalence of \tfn {} matrices}

Tensors $\cf$, $\widehat{\cf}$ are expressed by formulas \eqref{F}
and \eqref{Fhat} in particular bases of subalgebras ${\mathfrak g},
\wh{\mathfrak{g}}$ of Manin triples  $(\mathfrak{d},\mathfrak g ,
\wt{\mathfrak g})$ and $(\mathfrak{d},\wh{\mathfrak{g}},
\wb{\mathfrak{g}})$. However, both initial and dual/plural tensor do not depend on the choice of  bases in ${\mathfrak g}$ or
$\wh{\mathfrak{g}}$. Their geometric properties are thus independent as well.

Let us consider automorphisms of both Manin triples given by linear \tfn s of
$\mathfrak{g}$ and $\wh{\mathfrak{g}}$ that preserve their algebraic
structure. Let $A$ and $B$ be $d\times d$ matrices that transform
bases $T_a$ and $\widehat{T}_a$. Transformations  \eqref{C_mat} of
the form
\begin{equation}\label{automorphisms}
\left(\begin{array}{c}
T' \\ \widetilde T'
\end{array}\right)
=
\left(\begin{array}{cc}
A  & 0  \\
 0  & A^{-T}  \\
\end{array}\right)\cdot
\left(\begin{array}{c}
T \\ \widetilde T
\end{array}\right),
\quad
\left(\begin{array}{c}
\wh T' \\ \bar T'
\end{array}\right)
=
\left(\begin{array}{cc}
B & 0  \\
 0  & B^{-T}  \\
\end{array}\right)\cdot
\left(\begin{array}{c}
\wh T \\ \bar T
\end{array}\right)
\end{equation}
{then preserve the algebraic structure \eqref{commutation_on_d},
\eqref{commutation_on_d2} and duality \eqref{dual_bases} of bases
$(T,\widetilde T)$ and $(\wh T, \bar T)$.}

Transformations \eqref{automorphisms} induce changes in matrices
$E(s)$ and $\wh E(s)$ that are used in construction of background
tensors. We have
\begin{equation}
\nonumber
E'(s)={\cal A}\cdot E(s)\cdot {\cal {A}}^{T},\quad
 \widehat E'(s)={\cal B}\cdot \wh E(s)\cdot {\cal {B}}^{T}
\end{equation}
where
\begin{equation}
\nonumber
{\mathcal A} =\left(\begin{array}{cc}\unit & 0 \\ 0&A\end{array}\right), \qquad \mathcal{B} =\left(\begin{array}{cc}\unit & 0 \\ 0&B\end{array}\right).
\end{equation}
If the relation of bases in Manin triples is written as in
\eqref{pqrs}, then it is easy to check that
$$
\left(\begin{array}{c}
T' \\ \widetilde T'
\end{array}\right)
=
\left(\begin{array}{cc}
P'  & Q'  \\
 R'  & S'  \\
\end{array}\right)\cdot
\left(\begin{array}{c}
\wh T' \\ \bar T'
\end{array}\right)
$$
where
\begin{equation}\label{simplepqrs}
\left(\begin{array}{cc}
P'  & Q'  \\
 R'  & S'  \\
\end{array}\right)
=
\left(\begin{array}{cc}
APB\-1 & AQB^T  \\
A^{-T} RB\-1 & A^{-T}SB^T  \\
\end{array}\right)  .
\end{equation}
Therefore, matrices
\begin{equation}
\nonumber
\left(\begin{array}{cc}
P  & Q  \\
 R  & S  \\
\end{array}\right)
\ \rm{ and }\
\left(\begin{array}{cc}
P'  & Q'  \\
 R'  & S'  \\
\end{array}\right)
\end{equation}
define  transformations between \sm\  backgrounds  $\cf$ and  $\widehat{\cf}$ in various \coor s. That's why, from the perspective of  \pltp y, they can be considered equivalent.

\section{Sigma models with two-dimensional target space}\label{2dsm}

In this section we shall consider atomic \pltp y of \sm s whose target space is a two-dimensional solvable Lie group $\G$ with generators $T_1, T_2$ satisfying
\begin{equation}\label{solvable}
 [ T_1, T_2]= T_2.
\end{equation}
{The trace $f^k_{ik}$ of the structure constants is not zero since $f^2_{1 2}=1$ and it is known that this leads to mixed gauge and gravitational anomaly \cite{aagl} in the dual model. Yet, it is worth considering such groups in the context of integrable models \cite{hoatsey} and generalized supergravity \cite{hokico}.}
We parametrize the elements $g\in\G$ as $g=\e^{g_1T_1}\e^{g_2T_2}$. Since there are no spectators, the matrix $E(s)$ is constant. Choosing it in the form
\begin{equation}\nonumber
E(s)=\left(
\begin{array}{cc}
 \alpha  & \beta  \\
 \gamma  & 0  \\
\end{array}
\right),
\end{equation}
we find that the background tensor $\cf(g_1,g_2)$ calculated according to the formula \eqref{F} is given by
\begin{equation}\label{calF0}
\cf(g_1,g_2)=\left(
\begin{array}{cc}
1 & 0\\
 0  & \e^{g_1}  \\
\end{array}
\right)\cdot\left(
\begin{array}{cc}
 \alpha  & \beta  \\
 \gamma  & 0  \\
\end{array}
\right)\cdot\left(
\begin{array}{cc}
1 & 0\\
 0  & \e^{g_1}  \\
\end{array}
\right)=\left(
\begin{array}{cc}
 \alpha  & \e^{g_1} \beta  \\
\e^{g_1} \gamma  & 0 \\
\end{array}
\right).
\end{equation}
One can verify that $\cf$ is invariant with respect to $\G$ and its
symmetric part $\mathcal{G}$ is flat {metric. Since the target
manifold is two-dimensional, the torsion $H=d\mathcal{B}$ of all the
backgrounds discussed in this section vanishes. Therefore, the
$\mathcal{B}$-field can be eliminated by a gauge transformation and
the only relevant part of $\cf$ is the metric $\mathcal{G}$.}

\subsection{\pltp y}\label{pltpies}

In order to find \PL\ T-dual or plural models associated to \eqref{calF0}, we embed $\G$ into four-dimensional \dd\ $\D=(\G | \tG)$ with $\tG$ Abelian. The algebraic structure of four-dimensional \dd s was studied in \cite{hlasno:pltdm2dt}, where it was shown that for such \dd\ there are two nonequivalent Manin triples:
\begin{itemize}
\item Semi-Abelian triple $\mathfrak{d}=\mathfrak g \oplus \wt{\mathfrak g}$ with dual basis $(T_1,T_2,\wwt T^1,\wwt T^2)$ and Lie brackets (only nontrivial brackets are displayed)
\begin{equation}
[ T_1, T_2]= T_2, \qquad [T_1,\wwt T^2]=-\wwt T^2, \qquad [T_2, \wwt T^2]=\wwt T^1, \label{sabel}
\end{equation}
\end{itemize} or
\begin{itemize}
\item Type B non-Abelian triple $\mathfrak{d}=\wh{\mathfrak{g}} \oplus \wb{\mathfrak{g}}$ with dual basis $(\wh T_1,\wh T_2,\bar T^1,\bar T^2)$ and Lie brackets
\[
[\wh T_1,\wh T_2]=\wh T_2,\qquad  [\bar T^1,\bar T^2]=\bar T^1,
\]
\begin{equation}
[\wh T_1,\bar T^1]= \wh T_2,\qquad  [\wh T_1,\bar T^2]=-\wh T_1-\bar T^2, \qquad  [\wh T_2,\bar T^2]=\bar T^1. \label{typeb}
\end{equation}
\end{itemize}

Map relating both bases
\begin{align}
\wh T_1 & = -T_1+T_2, & \wh T_2 & = \wwt T^1+\wwt T^2, \nonumber\\
\bar T^1 & = \wwt T^2, & \bar T^2 & =  T_1 \label{2dplt:hs}
\end{align}
mentioned in the paper \cite{hlasno:pltdm2dt} preserves \eqref{dual_bases} and transforms Lie brackets \eqref{sabel} into \eqref{typeb}. However, there are two different classes of linear maps \eqref{C_mat} given by matrices
\begin{equation}\label{matC1}
C_1=\left(
\begin{array}{cccc}
 -1 & b_2 & b_2 b_3 & b_3 \\
 0 & 0 & b_1 b_2 & b_1 \\
 0 & 0 & b_1 b_2-1 & b_1 \\
 1 & \frac{1}{b_1}-b_2 & \left(\frac{1}{b_1}-b_2\right) b_3 & -b_3 \\
\end{array}
\right),\qquad b_1, b_2, b_3 \in \mathbb{R}, b_1\neq 0
\end{equation}
or
\begin{equation}\label{matC2}
C_2=\left(
\begin{array}{cccc}
 1 & b_2 & -b_2 b_3 & b_3 \\
 0 & b_1 & -b_1 b_3 & 0 \\
 0 & b_1 & 1-b_1 b_3 & 0 \\
 -1 & -b_2 & \frac{b_2 (b_1 b_3-1)}{b_1} & \frac{1}{b_1}-b_3 \\
\end{array}
\right),\qquad b_1, b_2, b_3 \in \mathbb{R}, b_1\neq 0
\end{equation}
that do the same and that define much richer set of decompositions\footnote{One can simplify these  matrices by \eqref{automorphisms}, \eqref{simplepqrs} and choose e.g. $b_1=1$, $b_2=b_3=0$ for $C_1$ and $b_2=0$ for $C_2$. To get the map \eqref{2dplt:hs} one has to choose $b_1=b_2=1$, $b_3=0$.}. Note that $\det C_1=-1$ and $\det C_2=1$. We will show that using these two maps to generate models ``plural'' to \eqref{Lagrangian} with $\cf$ given by \eqref{calF0} one gets substantially different models.

The first one, obtained from \eqref{Fhat}--\eqref{E0hat} and \eqref{matC1},
is given by background\footnote{We assume that elements of $\hG$ are
parametrized as $\hat{g}=\e^{\hat g_1 \wh T_1}\e^{\hat g_2 \wh
T_2}$.}
{\begin{equation}
\nonumber
\wh \cf(\hat g_1, \hat
g_2)=
\left(
\begin{array}{cc}
 \frac{e^{-\hat g_1} (b_3+\beta ) (b_3-\gamma )}{b_1 \left(\beta +\gamma +b_1 e^{\hat g_1} (\alpha -b_2 (\beta +\gamma ))\right)} & -\frac{-b_3-\beta -b_1 e^{\hat g_1} (\alpha -b_2 (\beta +\gamma ))}{\beta +\gamma +b_1 e^{\hat g_1} (\alpha -b_2 (\beta +\gamma ))} \\
 \frac{b_3-\gamma -b_1 e^{\hat g_1} (\alpha -b_2 (\beta +\gamma ))}{\beta +\gamma +b_1 e^{\hat g_1} (\alpha -b_2 (\beta +\gamma ))} & \frac{b_1 e^{\hat g_1}}{\beta +\gamma +b_1 e^{\hat g_1} (\alpha -b_2 (\beta +\gamma ))} \\
\end{array}
\right)
\end{equation}
with \emph{nonzero} ($b_1\neq 0$) \emph{scalar curvature}
$$\wh R =
-\frac{4 b_1 e^{\hat g_1}}{\beta +b_1 e^{\hat g_1} (\alpha -b_2
(\beta +\gamma ))+\gamma }.$$}
{The other one, obtained using
\eqref{matC2}, is given by}
\begin{equation}
\nonumber
\wh \cf(\hat g_1, \hat g_2)=
\left(
\begin{array}{cc}
 \frac{\alpha +b_2 (\beta +\gamma )}{\left(b_1 e^{\hat g_1} (b_3-\beta )-1\right) \left(b_1 e^{\hat g_1} (b_3+\gamma )-1\right)} & \frac{b_1 e^{\hat g_1} (b_3-\beta )}{b_1 e^{\hat g_1} (b_3-\beta )-1} \\
 -\frac{b_1 e^{\hat g_1} (b_3+\gamma )}{b_1 e^{\hat g_1} (b_3+\gamma )-1} & 0 \\
\end{array}
\right).
\end{equation}
This is a background with \emph{flat} metric,
so indeed, \emph{using \pltp y with two different maps \eqref{matC1} and \eqref{matC2}
we get two different \sm s.} This essential difference of curvature properties of the backgrounds remains true for any choice of $b_1, b_2, b_3$.

Similar results are obtained if we consider plural sigma models on $\bG$. In that case we use \tfn s between bases of semi-Abelian Manin triple and a ``dual'' to type-B Manin triple, i.e. the matrices \eqref{matC1} and \eqref{matC2} are multiplied from the left by the exchange matrix
\begin{equation}\label{bilinform}
D_0=\left(
\begin{array}{cccc}
 0 & 0 & 1 & 0 \\
 0 & 0 & 0 & 1 \\
 1 & 0 & 0 & 0 \\
 0 & 1 & 0 & 0 \\
\end{array}
\right).
\end{equation}

\subsection{\pltd y}\label{pltdies}

Maybe surprisingly, we observe the same phenomenon for \pltd y
as well. Dual sigma models can be obtained by exchange of Manin
triples {$(\cd, \cg, \tcg)$ and $(\cd, \tcg, \cg)$} mediated by the
matrix \eqref{bilinform}. On the other hand, there are more general
maps between bases of the semi-Abelian Manin triple and its dual.
Linear maps on $\cd$ that switch the roles of $T_a$ and $\widetilde
T^{a}$ in \eqref{sabel} and meanwhile preserve the duality of the
bases \eqref{dual_bases} are given by automorphisms \eqref{C_mat}
where the matrix $C$ is either
\begin{equation}\label{matD1}
D_1=\left(
\begin{array}{cccc}
 0 & 0 & 1 & 0 \\
 0 & 0 & -\frac{b_2}{b_1} & \frac{1}{b_1} \\
 1 & b_2 & -b_2 b_3 & b_3 \\
 0 & b_1 & -b_1 b_3 & 0 \\
\end{array}
\right), \qquad b_1, b_2, b_3 \in \mathbb{R}, b_1\neq 0
\end{equation}
or
\begin{equation}\label{matD2}
D_2=\left(
\begin{array}{cccc}
 0 & 0 & -1 & 0 \\
 0 & b_1 & b_1 b_3 & 0 \\
 -1 & b_2 & b_2 b_3 & b_3 \\
 0 & 0 & \frac{b_2}{b_1} & \frac{1}{b_1} \\
\end{array}
\right), \qquad b_1, b_2, b_3 \in \mathbb{R}, b_1\neq 0
\end{equation}
with determinants of $D_1$ and $D_2$ equal to $\pm 1$. Note that
$D_1$ is equal to the exchange matrix \eqref{bilinform} for
$b_1=1,b_2=b_3=0$.
Inserting each of these matrices into \eqref{Fhat}--\eqref{E0hat} one gets again two different ``dual''  models.

The first model, obtained using \eqref{matD1} 
and parametrization $\tilde{g}=\e^{\tilde g_1 \widetilde
T^1}\e^{\tilde g_2 \widetilde T^2}$, has \emph{flat} background
\begin{equation}
\label{dualF2} \widetilde \cf(\tilde g_1, \tilde g_2)= \left(
\begin{array}{cc}
 0 & \frac{1}{b_1 (b_3+\gamma )+\tilde{g}_2} \\
 \frac{1}{b_1 (\beta -b_3)-\tilde{g}_2} & \frac{\alpha +b_2 (\beta +\gamma )}{(b_1 (b_3-\beta )+\tilde{g}_2) (b_1 (b_3+\gamma )+\tilde{g}_2)} \\
\end{array}
\right),
\end{equation}
while background obtained using \eqref{matD2} is given by tensor
\begin{equation}\label{dualF3}
\widetilde \cf(\tilde g_1, \tilde g_2)= \left(
\begin{array}{cc}
 \frac{1}{\alpha -(\beta +\gamma ) (b_2+b_1 \tilde{g}_2)} & \frac{b_1 (b_3+\beta )}{(\beta +\gamma ) (b_2+b_1 \tilde{g}_2)-\alpha } \\
 \frac{b_1 (b_3-\gamma )}{(\beta +\gamma ) (b_2+b_1 \tilde{g}_2)-\alpha } & -\frac{b_1^2 (b_3+\beta ) (b_3-\gamma )}{(\beta +\gamma ) (b_2+b_1 \tilde{g}_2)-\alpha } \\
\end{array}
\right)
\end{equation} \emph{with nonzero scalar curvature}
$$\widetilde R=\frac{4}{(\beta +\gamma ) (b_1 \tilde{g}_2+b_2)-\alpha }.$$
One can see that once more we get two different \pl\ T-dual \sm s no
matter what the parameters $b_1, b_2$ and $b_3$ are.

It is well known that (non-)Abelian T-duality is induced by
matrix $D_0$ that is a special case of $D_1$. 
{In fact, dualizing $\cf$ using $D_0$ we get tensor
\begin{equation}
\nonumber
\wwt \cf'(\tilde g_1, \tilde g_2) = \left(
\begin{array}{cc}
 0 & \frac{1}{\gamma +\tilde{g}_2} \\
 \frac{1}{\beta -\tilde{g}_2} & -\frac{\alpha }{(\beta -\tilde{g}_2) (\gamma +\tilde{g}_2)} \\
\end{array}
\right)
\end{equation}
that can be brought to the form \eqref{dualF2} by coordinate transformation
\begin{equation}
\nonumber
\tilde{g}_1 = \tilde{g}'_1-\frac{b_2 \tilde g'_2}{b_1},\qquad \tilde{g}_2=\frac{\tilde{g}'_2}{b_1}+b_3.
\end{equation}
}
Alternatively, it can be obtained by gauging the (non-)Abelian isometry and
introduction of Abelian Lagrange multipliers. {We shall investigate whether} the duality induced by matrix $D_2$ can be obtained in a similar
way.

Matrix $D_2$
can be transformed by automorpisms \eqref{automorphisms}, \eqref{simplepqrs} to the form
\begin{equation}\label{matD20}
D'_2=\left(
\begin{array}{cccc}
 0 & 0 & -1 & 0 \\
 0 &1 & 0 & 0 \\
 -1 & 0 & 0 & 0 \\
 0 & 0 &0 & 1 \\
\end{array}
\right).
\end{equation}
Up to the change of sign necessary for being an automorphism of {semi-Abelian Manin triple}, matrix $D'_2$ acts by switching $T_1\leftrightarrow \wwt T_1$.
One may suspect that this can actually be Buscher duality with respect to one-dimensional subgroup\footnote{Dualities of this form are called factorised in \cite{LuOst}.} of isometry group 
$\G$. This subgroup is generated by left-invariant vector field $V_1=\partial_{g_1}-g_2\,\partial_{g_2} $ that together with $V_2=\partial_{g_2} $ satisfies \eqref{solvable}.

{To check our suspicion we have to find adapted coordinates $\{s_1,h_1\}$ such that $V_1$ becomes $V'_1=\partial_{h_1}$ and $\cf$ becomes independent of $h_1$. Suitable transformation of coordinates is given by
\begin{equation}
\label{gtoY} g_1= h_1,\qquad g_2=s_1\,\e^{-h_1}.
\end{equation}
Tensor \eqref{calF0} is then transformed to the form
\begin{equation}\label{buscherdual}
\cf'(s_1,h_1) = \left(
\begin{array}{cc}
 0 & \gamma  \\
 \beta  & \alpha -(\beta +\gamma ) s_1 \\
\end{array}
\right).
\end{equation}
Treating $s_1$ as spectator we may dualize $\cf'$ with respect to $h_1$. Buscher rules that follow from \eqref{Fhat}--\eqref{E0hat} give tensor
\begin{equation}\label{buscherdualF}
\widetilde \cf(s_1, \tilde h_1) =
\left(
\begin{array}{cc}
 \frac{\beta  \gamma }{(\beta +\gamma ) s_1-\alpha } & \frac{\gamma }{(\beta +\gamma ) s_1-\alpha } \\
 \frac{\beta }{\alpha -(\beta +\gamma ) s_1} & \frac{1}{\alpha -(\beta +\gamma ) s_1} \\
\end{array}
\right)
\end{equation}
and subsequent change of coordinates
\begin{equation}
\nonumber
s_1 = b_1 \tilde{g}_2+b_2,\qquad h_1=b_1 b_3 \tilde{g}_2-\tilde{g}_1
\end{equation}
restores \eqref{dualF3}.
}

{Alternative formulation of duality given by matrix
\eqref{matD20} follows from gauge invariant parent action
\begin{align*}
S[h_1,s_1,A_-,A_+,\tilde{h}_1]=\frac{1}{2\pi}\int D_- h_1\left(\alpha -{s_1} (\beta +\gamma )\right)D_+ h_1+\\
+D_- h_1\,\beta\, \partial_+ s_1 + \partial_-
s_1\,\gamma\,D_+ h_1 +\tilde h_1 (\partial_- A_+
-\partial_+ A_-)
\end{align*}
where
$$D_{\pm}h_1=\partial_{\pm}h_1
+A_{\pm}h_1.$$
Integrating out gauge fields $A_+$ and $A_-$
we obtain \sm\ with background tensor \eqref{buscherdualF} that can be brought to the form \eqref{dualF3} by coordinate transformation.}

{One may also ask why Buscher duality with respect to $V_2$ is not included in \eqref{matD1} or \eqref{matD2}. The reason is that change of bases
$$ T'_1 = T_1,\quad T'_2=\pm\wwt T_2,\quad \wwt T'_1=\wwt T_1,\quad \wwt T'_2=\pm T_2$$
is not an automorphism of the Manin triple given by \eqref{solvable} and $[\wwt T_1,\wwt T_2]=0$.}

\section{Sigma models with four-dimensional target space}\label{4dsm}

One can also ask if there are some ``\pl\ identities'' preserving the structure of semi-Abelian Manin triple, i.e. \pltp ies generated by automorphisms of $\mathfrak{d}=\mathfrak g \oplus \wt{\mathfrak g}$ that preserve both Lie brackets \eqref{sabel} and duality of the basis \eqref{dual_bases}. The answer is positive, and the mappings can have two possible forms given by matrices
\begin{equation}\label{matI1}
I_1=\left(
\begin{array}{cccc}
 1 & b_2 & -b_2 b_3 & b_3 \\
 0 & b_1 & -b_1 b_3 & 0 \\
 0 & 0 & 1 & 0 \\
 0 & 0 & -\frac{b_2}{b_1} & \frac{1}{b_1} \\
\end{array}
\right), \qquad b_1, b_2, b_3 \in \mathbb{R}, b_1 \neq 0
\end{equation} and
\begin{equation}\label{matI2}
I_2=\left(
\begin{array}{cccc}
 -1 & b_2 & b_2 b_3 & b_3 \\
 0 & 0 & \frac{b_2}{b_1} & \frac{1}{b_1} \\
 0 & 0 & -1 & 0 \\
 0 & b_1 & b_1 b_3 & 0 \\
\end{array}
\right) \qquad b_1, b_2, b_3 \in \mathbb{R}, b_1 \neq 0.
\end{equation}
Using \eqref{matI1} in atomic \pltp y transformation of the model given by \eqref{calF0} we get a sigma model in flat background. The example, however, is not particularly illuminating since the condition \eqref{cond_plural} is not satisfied for \eqref{matI2} and plural background cannot be calculated. For further investigation we focus on sigma models in four-dimensional target space and introduce spectator fields.

In the papers \cite{php}, \cite{hpp} \pltd y with spectators was used to study non-Abelian T-duals of \sm s in flat Minkowski space. {Given the metric $\eta=\text{diag}(-1,1,1,1)$ in coordinates $\{t,x,y,z\}$, we consider Killing vectors
$$T_1 := K_3= z\partial_t+t \partial_z, \qquad T_2 := L_2 + K_1 = x\partial_t+(t+z)\partial_x-x\partial_z$$ satisfying $[ T_1, T_2]= T_2$. These vectors generate a solvable two-dimensional group $\G$ of symmetries of the background $\eta$. The coordinates $\{s_1,s_2,g_1,g_2\}$ given by\footnote{The action of $\G$ is not free and transitive for $t+z=0$, i.e. for $s_1=0$. We have to restrict our calculations to coordinate patches with $s_1\neq 0$.}
\begin{align}
\nonumber t&=\frac{1}{2} e^{-g_1} \sqrt{\left| s_1\right| } \left(\text{sgn}(s_1)+e^{2 g_1} \left(g_2^2+1\right)\right),\\ \nonumber x&=-e^{g_1} g_2 \sqrt{\left| s_1\right| },\\ \label{ad_coord} y&=s_2,\\ \nonumber z&=-\frac{1}{2} e^{-g_1} \sqrt{\left| s_1\right| } \left(\text{sgn}(s_1)+e^{2 g_1} \left(g_2^2-1\right)\right)
\end{align}
are adapted to the action of $\G$. After the transformation of coordinates \eqref{ad_coord} the flat metric acquires the simple form
\begin{equation}\label{Minkowski}
\cf(s_1,s_2,g_1,g_2)=\left(
\begin{array}{cccc}
 -\frac{1}{4 s_1} & 0 & 0 & 0 \\
 0 & 1 & 0 & 0 \\
0 & 0 &  s_1 &
   0 \\
 0 & 0 &0& e^{2 g_1} s_1
\end{array}
\right)
\end{equation}
from which one gets $E(s)$ by setting $g_1=0$.
The group $\G$ can be embedded into semi-Abelian \dd\ with algebraic structure
\eqref{sabel} allowing to find dual/plural backgrounds\footnote{For further {details concerning the process of finding the adapted coordinates see \cite{php} where this particular case was denoted $S_{2,20}$.}}.}

Inserting matrices \eqref{matI1} and \eqref{matI2} into \eqref{Fhat}--\eqref{E0hat} one gets two different \sm s on $\N \times \G$ related to the original model in background \eqref{Minkowski} by \pltp y. The first, obtained using \eqref{matI1}, is given by background
\begin{equation}\label{dualI1}
 \cf(s_1,s_2,g_1,g_2)=\left(
\begin{array}{cccc}
 -\frac{1}{4 s_1} & 0 & 0 & 0 \\
 0 & 1 & 0 & 0 \\
 0 & 0 & \left(b_2^2+1\right) s_1 &
   b_1 e^{g_1} (b_2
   s_1-b_3) \\
 0 & 0 & b_1 e^{g_1} (b_3+b_2
   s_1) & b_1^2 e^{2 g_1} s_1
\end{array}
\right)
\end{equation}
that is \emph{flat} and \emph{torsionless} for any values\footnote{Note that \eqref{matI1} reduces to identity matrix for $b_1=1,b_2=b_3=0$. Consequently, \eqref{dualI1} reduces to \eqref{Minkowski}.} of $b_1, b_2, b_3$. The second background, obtained using \eqref{matI2}, is given by tensor
\begin{equation}\label{dualI2}
 \cf(s_1,s_2,g_1,g_2)= \left(
\begin{array}{cccc}
 -\frac{1}{4 s_1} & 0 & 0 & 0 \\
 0 & 1 & 0 & 0 \\
 0 & 0 & \frac{b_3^2}{s_1}+s_1 &
   \frac{e^{g_1} (b_3-b_2
   s_1)}{b_1 s_1} \\
 0 & 0 & \frac{e^{g_1} (b_3+b_2
   s_1)}{b_1 s_1} & \frac{e^{2
   g_1}}{b_1^2 s_1}
\end{array}
\right).
\end{equation}
Similarly to the previous case, the torsion vanishes. However, the symmetric part of \eqref{dualI2}, i.e. the metric, has vanishing scalar curvature but \emph{nontrivial Ricci tensor}.
Using transformation of coordinates
\begin{align*}
s_1 &= \frac{3 u^4}{b_1^2}+\frac{6 u^2 z_3}{b_1}-2 u v+z_3^2, &  s_2 &= z_4,\\ g_1 &= \frac{1}{2} \ln \left(\frac{3 u^2}{b_1^2}+\frac{6 z_3}{b_1}-\frac{2 v}{u}+\frac{z_3^2}{u^2}\right),&g_2 &= \frac{b_1 b_3}{\sqrt{\frac{3 u^2}{b_1^2}+\frac{6 z_3}{b_1}+\frac{z_3^2-2 u v}{u^2}}}-b_1 u z_3-u^3
\end{align*}
one can bring this background to a \emph{pp-wave} in the Brinkmann form
$$ds^2= 2\frac{z_3^2}{u^2} du^2 + 2 du dv +dz_3^2 + dz_4^2$$
{identified in \cite{php,hpp} to be one of the gauged WZW models considered in \cite{tsey:revExSol, tseyt94}.}
We see that we again get two different \sm s, this time produced by \pltp ies that do not change the Manin triple.

\section{\pltd ies and pluralities generated by \PL {} identities}\label{PLbyPLids}

Comparing \tfn\ matrices \eqref{matD1}, \eqref{matD2} that generate \pltd ies and \eqref{matI1}, \eqref{matI2} that generate \PL {} identities one may notice that they are related by canonical duality matrix \eqref{bilinform} that exchanges generators $T_a$ and $\widetilde T^{a}$. Indeed, it is easy to check that
\begin{equation} \label{pltdrelns}
D_1=D_0\cdot I_1,\quad D_2=D_0\cdot I_2.
\end{equation}
That means that all \pltd ies described in subsection \ref{pltdies} can be obtained by canonical \natd y of \sm s generated by \PL\ identities of the initial model. {Writing \eqref{C_mat} as
\begin{equation}\nonumber
\begin{pmatrix}\widetilde T \\
T \end{pmatrix} = D_0 \cdot \begin{pmatrix}T \\ \widetilde T \end{pmatrix} = D_0 \cdot I \cdot \begin{pmatrix}T \\ \widetilde T \end{pmatrix}
\end{equation}
we verify that this holds not only for the four-dimensional semi-Abelian \dd, but for general \dd\ $(\G | \tG)$.}

One can try to obtain similar relation for \pltp ies described in subsection \ref{pltpies}. However, question then is what is the ``canonical \pltp y''. Motivated by \eqn s \eqref{pltdrelns} we can define it for Manin triples $(\mathfrak{d},\mathfrak g, \wt{\mathfrak g})$ and $(\cd,\hat{\mathfrak g},\bar{\mathfrak{g}})$ with \eqref{sabel} and \eqref{typeb} by
\begin{equation}\label{C0}
C_0=\left(
\begin{array}{cccc}
 -1 & 0 & 0 & 0 \\
 0 & 0 & 0 & 1 \\
 0 & 0 & -1 & 1 \\
 1 & 1 & 0 & 0 \\
\end{array}
\right),
\end{equation}
i.e. by the matrix $C_1$ with $b_1=1,b_2=b_3=0$. {Relabeling $b_1\to
\frac{1}{b_1},b_2\to -b_2,b_3\to -b_3$ in $I_1$ and $b_2\to
-b_2,b_3\to -b_3$ in $I_2$ we then have}
\begin{equation}
\nonumber
C_1=C_0\cdot I_1,\quad C_2=C_0\cdot I_2
\end{equation}
and all \pltp ies from subsection \ref{pltpies} can be obtained by
{\pltp y of \sm s generated by \eqref{C0} and \PL\ identities of the
initial model.} Unfortunately, the choice \eqref{C0} is by far not
unique, and, as opposed to duality, we can hardly call it
``canonical'' \pltp y.

We have obtained similar results for \pltp ies {of \sm s embedded in six-dimensional \dd s.}

\section{Conclusion}

The examples presented in sections \ref{2dsm} and \ref{4dsm} prove that families of \sm s related by \pltp y may depend not only on the algebraic structure of Manin triple 
but also on the way how the given Manin triple is embedded in the \dd.

This holds for the \pltd y as well, so we may ask what should be considered as ``true'' T-duality. Is it only the procedure introduced in \cite{buscher:ssbfe,delaossa:1992vc, rocekver:duality} that uses gauging of initial \sm {} and is alternatively described in \cite{klise} as an exchange of dual bases
$T_a$ and $\widetilde T^{a}$ of the isotropic subalgebras of  Manin triple?
Or can we admit any linear \tfn\ of bases of the \dd\ that give decompositions isomorphic to the Manin triple obtained by the exchange?

Possible answer to the question above follows from expressions \eqref{F}
and \eqref{Fhat} of both initial and dual/plural
tensor. Namely, from these expressions we can see
that tensors $\cf$ and $ \wh\cf$ depend both on algebraic structure
of Manin triple and bilinear forms $E(s), \wh E(s)$. \PL{} identity
does not change the initial Manin triple (only its embedding in the
\dd) but changes the bilinear form $E(s)$ according to the
formula \eqref{E0hat}. Subsequent canonical duality or plurality
then changes the Manin triple and produce \PL\ T-dual/plural tensor.
Moreover, it turns out that non-canonical dualities may be hidden canonical dualities with respect to subgroups of the initial group of isometries.



\begin{thebibliography}{99}

\bibitem{buscher:ssbfe} T. H. Buscher, \emph{A symmetry of the String Background Field Equations}, {Phys. Lett. B} 194 (1987) 51.

\bibitem{delaossa:1992vc}
X. C. de~la Ossa and F.~Quevedo, \emph{Duality symmetries from non-abelian   isometries in string theories},  Nucl. Phys. {B 403} (1993) 377, [hep-th/9210021].

\bibitem{rocekver:duality}
M. Ro\v cek and E. Verlinde, \emph{Duality, Quotients, and Currents}, {Nucl. Phys. B 373} (1992) 630, [hep-th/9110053].

\bibitem{1012.1320 sfethomp}  K. Sfetsos and D. C. Thompson, \emph{On non-abelian T-dual geometries with Ramond fluxes}, Nucl. Phys.  B 846 (2011)  21, [arXiv:1012.1320].

\bibitem{1104.5196LCST}  Y. Lozano, E. \'O Colg\'ain, K. Sfetsos, and D. C. Thompson, \emph{Non-abelian T-duality, Ramond fields and coset geometries}, JHEP 06 (2011) 106, [arXiv:1104.5196].



\bibitem{KlebWitt} G. Itsios, Y. Lozano, J. Montero, and C. N{\'u}{\~{n}}ez, \emph{The $AdS_5$ non-Abelian T-dual of Klebanov-Witten as a $\mathcal{N}=1$ linear quiver from M5-branes}, JHEP 09 (2017) 38, [arXiv:1705.09661]

\bibitem{INNSZ} G. Itsios, H. Nastase, C. N{\'u}{\~{n}}ez, K. Sfetsos, and S. Zacarias, \emph{Penrose limits of Abelian and non-Abelian T-duals of $AdS_5\times S^5$ and their field theory duals}, JHEP 01 (2018) 71 (2018), [arXiv:1711.09911].

\bibitem{klise}{C. Klim\v c\'ik and P. \v Severa, \emph{Dual non-Abelian duality and the Drinfeld double}, Phys. Lett. B 351 (1995) 455, [hep-th/9502122].}

\bibitem{klim:proc}{C. Klim\v c\'ik, \emph{Poisson--Lie T-duality}, Nucl. Phys. Proc. Suppl. 46 (1996) 116, [hep-th/9509095].}

\bibitem{Bouwknegt2017}
{P. Bouwknegt, M. Bugden, C. Klim{\v{c}}{\'i}k, and K. Wright, \emph{Hidden isometry of ``T-duality without isometry''}, JHEP 08 (2017) 116, [arXiv:1705.09254].}

\bibitem{BorWulff1} R. Borsato and L. Wulff, \emph{Integrable deformations of T-dual {$\sigma$}-models}, Phys. Rev. Lett. 117 (2016) 251602, [arXiv:1609.09834].

\bibitem{BorWulff} R. Borsato and L. Wulff, \emph{On non-abelian T-duality and deformations of supercoset string sigma-models}, JHEP 10 (2017) 24, [arXiv:1706.10169].

\bibitem{HoaSei} B. Hoare and F. K. Seibold, \emph{Poisson-Lie duals of the $\eta$ deformed symmetric space sigma model}, JHEP 11 (2017) 14, [arXiv:1709.01448].

\bibitem{unge:pltp} R. von Unge, \emph{Poisson--Lie T--plurality}, JHEP 07 (2002) 014, [hep-th/0205245].

\bibitem{hlasno:3dsm2}
L. Hlavat\'y and L. \v Snobl, \emph{Poisson--Lie T--plurality of three-dimensional conformally invariant sigma models II : Nondiagonal metrics and dilaton puzzle}, JHEP 10 (2004) 045, [hep-th/0408126].

\bibitem{gom:ctd}
X. Gomez, \emph{Classification of three--dimensional Lie bialgebras}, J. Math. Phys. 41 (2000) 4939.

\bibitem{hlasno:pltdm2dt}
L.~Hlavat\'y and L.~\v{S}nobl, \emph{Classification of Poisson--Lie T-dual models with two-dimensional targets}, Mod. Phys. Lett. A 17 (2002) 429, [hep-th/0110139].

\bibitem{snohla:ddoubles}
L.~\v{S}nobl and L.~Hlavat\'y, \emph{Classification of 6-dimensional real Drinfel'd doubles}, Int. J. Mod. Phys. A 17 (2002) 4043, [math.QA/0202209].

\bibitem{hlapevoj}
L.~Hlavat\'y, I.~Petr, and V. \v St\v ep\'an,
\emph{Poisson--Lie T-plurality with spectators}, J. Math. Phys. 50 (2009) 043504.

\bibitem{hlafilp:uniq}L. Hlavat\'y and F. Petr\'asek, \emph{On
uniqueness of T-duality with spectators}, Int. J. Mod. Phys. A  31 (2016) 1650143, [arXiv:1606.02522].

\bibitem{hlapet:cqgrav} L. Hlavat\'y and I. Petr, \emph{Plane-parallel waves as duals of the flat background}, Class. Quantum Grav.  32 (2015) 035005,  [arXiv:1406.0971].

\bibitem{php} F. Petr\'asek, L. Hlavat\'y, and I. Petr, \emph{Plane-parallel waves as duals of the flat background II: T-duality with spectators}, Class. Quantum Grav. 34 (2017) 155003, [arXiv:1612.08015].

\bibitem{hpp} L. Hlavat\'y, I. Petr, and F. Petr\'asek, \emph{Plane-parallel waves as duals of the flat background III: T-duality with torsionless $B$-field}, Class. Quantum Grav. 35 (2018) 075012, [arXiv:1711.08688].

\bibitem{PWZ} J. Patera, R. T. Sharp, P. Winternitz, and H. Zassenhauss, \emph{Subgroups of the Poincar\'e group and their invariants}, J. Math. Phys. 17 (1976) 977.

\bibitem{LuOst} D. L\"ust and D. Osten, \emph{Generalised fluxes, Yang-Baxter deformations and the O(d,d) structure of non-abelian T-duality}, JHEP 05 (2018) 165, [arXiv:1803.03971].

\bibitem{aagl} E. \'Alvarez, L. \'Alvarez-Gaum\'e, Y. Lozano, \emph{On non-abelian duality}, Nucl. Phys. B 424 (1994) 155, [hep-th/9403155v4].

\bibitem{hoatsey} B. Hoare and A. A. Tseytlin, \emph{Homogeneous Yang-Baxter deformations as non-abelian duals of the $AdS_5$ sigma-model}, J. Phys. A 49 (2016) 494001, [arXiv:1609.02550v3]

\bibitem{hokico}  M. Honga, Y. Kima, and E. \'O Colg\'ain, \emph{On non-Abelian T-duality for non-semisimple groups}, Eur. Phys. J. C 78 (2018) 1025 [arXiv:1801.09567].

\bibitem{tsey:revExSol}{A. A. Tseytlin, \emph {Exact solutions of closed string theory}, Class. Quantum Grav. 12 (1995) 2365, [hep-th/9505052].}

\bibitem{tseyt94} A. A. Tseytlin, \emph{Exact string solutions and duality}, in proceedings of the 2nd Journee Cosmologique within the framework of the International School of Astrophysics, D. Chalonge, Paris, France, 2-4 June 1994, pp. 371-398, [hep-th/9407099].
\end{thebibliography}
\end{document}